\newcommand{\bra}[1]{\langle #1|}
\newcommand{\ket}[1]{|#1\rangle}

\documentclass[aps,prl,twocolumn,groupedaddress]{revtex4}
\usepackage{graphicx}
\usepackage{bm}
\usepackage{color}
\begin{document}

%\preprint{}

%Title of paper$
\title{The Impact of an Oxygen Dopant in an ideal \protect{Bi$_2$Sr$_2$CaCu$_2$O$_{8+\delta}$}
 Crystal}


\author{S. Johnston$^{1,2}$}
%\email[]{steve@lorax.uwaterloo.ca}
\author{F. Vernay$^3$}
\author{T.P. Devereaux$^{2}$}
\affiliation{$^1$Department of Physics, University of Waterloo, Waterloo, Ontario, N2L 3G1, Canada}
\affiliation{$^2$Department of Photon Science, Stanford Linear Accelerator
  Center, Stanford University, Menlo Park, CA, 94025, USA}
\affiliation{$^3$Paul Scherrer Institut, Condensed Matter Theory Group,
  Villigen PSI, Switzerland}

\begin{abstract}
Recent scanning tunneling microscopy studies have shown that local nanoscale pairing 
inhomogenities are correlated with interstitial oxygen dopants in Bi$_2$Sr$_2$CaCu$_2$O$_{8+\delta}$.  
Combining electrostatic and cluster calculations, 
in this paper the impact of a dopant on the local Madelung 
and charge transfer energies, magnetic exchange $J$, Zhang-Rice mobility, and 
interactions with the lattice is investigated. It is found that
electrostatic modifications locally increases the charge transfer energy
and slightly suppresses $J$. It is further shown that coupling to c-axis phonons is strongly modified
near the dopant. The combined effects of electrostatic modifications and
coupling to the lattice yield broadened spectral features, reduced charge gap
energies, and a sizable local increase of $J$. 
This implies a strong local interplay between antiferromagnetism, polarons, and superconducting pairing. 
\end{abstract}

\date{\today}
\maketitle

The precise role of the atoms lying off the CuO$_2$ planes has been an intriguing puzzle 
in the cuprates. While largely thought to provide a charge reservoir
to dope holes into the CuO$_2$ plane, it has become clear that an 
understanding of the pairing mechanism will require addressing the large
variations in T$_c$ arising from the local environment surrounding the CuO$_2$
planes\cite{Eisaki}. 
Empirically, the role of the apical or axial orbitals has been a vehicle linking
T$_c$ to either an effective electron hopping $t^{\prime}$ along diagonal
Cu-Cu bonds\cite{Pavarini}, or
to the stability of the Zhang-Rice singlet (ZRS)\cite{OhtaPRB}.  
However to date these arguments have pointed out possible links but offer little microscopic 
reason for the impact on T$_c$ itself.  

Scanning tunneling microscopy (STM) 
in Bi$_2$Sr$_2$CaCu$_2$O$_{8+\delta}$ (Bi2212) has revealed that nanoscopic inhomogeneity is correlated with
the location of interstitial oxygen dopant atoms\cite{McElroy} or the superlattice modulation\cite{superlattice}. 
The location of dopants has been correlated with
suppressed peak features with larger gap energies
in the observed local density of states (LDOS), and have been associated with local modification of superconducting pairing\cite{Nunner}.
This suggests a non-trivial link between the dopant atoms and the electronic properties of the material on a local level.

In metallic systems, such defects can be effectively screened and have little impact on the electrostatics of 
the material.  This is in contrast to the cuprates, which have poor screening along the $c$-axis 
and are unable to effectively screen the dopant's impurity charge.  As a result, these dopants 
and accompanying strutural changes may have a substantial impact on the
electrostatic properties of the 
material.  This may be reflected in quantities such as the charge transfer
energy $\Delta$, effective hoppings $t, t^{\prime}$, magnetic exchange 
interaction $J$, or electron-phonon (el-ph) coupling strength $\lambda$. 

From $t-J$ studies\cite{Maska}, it has been argued that dopants give
an enhancement of $J$ and thus larger gap energies, yet the overall shape of
the LDOS suggests that incoherence, giving 
broad spectral features, is an important ingredient
to understand local pairing modifications. 

In order to quantitatively 
address these issues electrostatic Ewald calculations are performed for Bi2212
%Bi$_2$Sr$_2$CaCu$_2$O$_{8+\delta}$ (Bi2212) 
supercells to determine the spatial dependence of the Madelung energies around
atomic sites in the crystal. It is found that while Madelung energies on O and Cu are spatially varied on the scale of eVs, 
these changes largely cancel and $\Delta$ is slightly increased near the dopant, yielding an overall suppression locally to $J$.
This information is then combined into exact diagonalization (ED) cluster studies,
yielding effective parameters $t, ~t^{\prime},~J$ and $J^{\prime}$. 
Large $O(1)$ changes are found both in $t^{\prime}$ as well as $c-$axis
el-ph coupling $\lambda$, quantities which are known to strongly modify a $d_{x^{2}-y^{2}}$ pairing interaction. 
Finally ED cluster studies including c-axis phonons are shown to produce
broadened spectra, a reduced charge gap, in agreement with
experiments\cite{McElroy}. As a consequence, a sizable local increase of $J$
results due to reduced gap via a gain in lattice energy, implying a strong
interplay between el-ph coupling, and local superconducting pairing and antiferromagnetic correlations.

\begin{figure*}[t]
 \includegraphics[width=\textwidth]{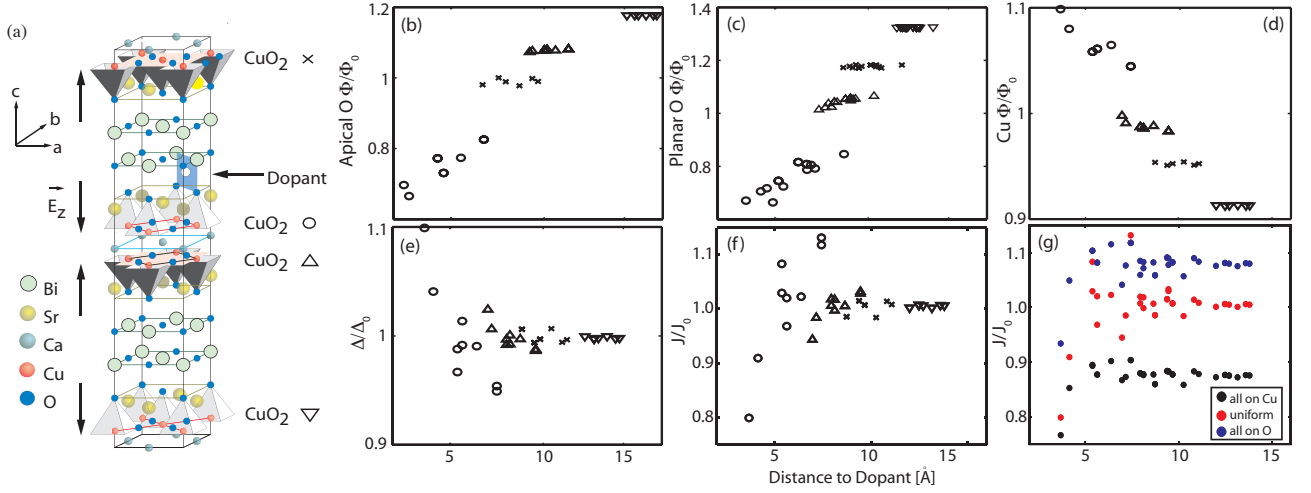}
 \caption{\label{Fig:1} (a) Schematic location of the dopant oxygen in Bi-2212 unit cell.  
Individual CuO$_2$ planes are labeled by the symbols shown.  The arrows on the left 
indicate the orientation of the local oxygen crystal fields for the undoped lattice.  (b - d) 
 Madelung energies of the apical O, planar O and Cu sites, respectively, in the doped 
 lattice.  The local value of the charge transfer energy $\Delta$ (e) and derived
 exchange interaction $J$ (f), obtained by averaging $\Delta\Phi_M$ between 
 Cu and its four neighbouring O sites.  (g) Resulting $J$ for
 different distributions of charges. All values have been 
 normalized to the undoped lattice values.  The distance to dopant 
 is defined as the distance to the closest dopant, accounting for the periodicity 
 of the superlattice.
 }
\end{figure*}

Electrostatic calculations using Ewald's method are performed on
a supercell consisting of $3\times3\times1$ Bi-2212 unit cells and containing 270 atoms.  
We have examined supercells up to 
four times this size and found that the results are not qualitatively different.   Each 
unit cell contains 2-primitive cells stacked along the $c$-axis for a total of 4 CuO$_2$ 
planes, as shown in Fig. \ref{Fig:1}a. Using formal valences for the atoms along with the known
structural data\cite{KovalevaPRB2004}, the Madelung energies $\Phi$
obtained are $\Phi_{apex,plane} = 18.48, 10.16$ eV for the apical, 
planar oxygen sites respectively, and $\Phi_{Cu} = -38.22$ eV for the copper site, 
consistent with values reported previously\cite{OhtaPRB}. The
charge transfer $\Delta$ is related to the difference in Madelung energies for the Cu and O sites, 
$\Delta\Phi_M = \Phi_O - \Phi_{Cu}$ and is given by:
\begin{equation}
\Delta = \frac{\Delta\Phi_M}{\epsilon(\infty)} - I_{Cu}(2) + A_O(2) - \frac{e^2}{d_p}
\label{Eq:1}
\end{equation}
where $I_{Cu}(2)$ and $A_O(2)$ are the second ionization and electron affinity energies 
for the Cu and O sites respectively.  The factor of $e^2/d$ represents the contribution 
of the Coulomb interaction between the introduced electron-hole pair.  In this
work we take the dielectric constant $\epsilon(\infty) = $ 3.5 and 
$I_{Cu}(2) - A_O(2) + e^2/d = $ 10.9 eV\cite{OhtaPRB}, yielding
$\Delta = 2.92$ eV. Besides setting the energy scale for gap  excitations,
$\Delta$ largely governs the magnetic exchange energy $J$. 
Using a canonical standard set of parameters in the limit of small hole hopping\cite{params,EskesPRB1993},
an
exchange energy $J \sim 300$ meV is obtained. This value larger than that found in
experiments\cite{TPDRH} results from the limitations of a perturbative
expression for experimentally relevant parameter sets.

The Ewald calculation was then repeated with a single oxygen dopant atom inserted into the unit
cell, shown schematically in Fig. \ref{Fig:1}a,
and the neighbouring atoms displaced as indicated from recent LDA studies\cite{HePRL2006}. 
The oxygen dopant was assigned formal valence, with surplus charge 
distributed equally among orbitals in the closest CuO$_2$ planes.  

The site-dependent Madelung energies are presented in Fig. \ref{Fig:1}b-d, showing
large scale variations for sites closest to the dopant atom.
Suppressions/enhancements
of O/Cu Madelung energies are observed, respectively,
rising/falling to bulk values, shifted by the presence of doped holes, further away from 
the dopant. However, since the
relative sign of the Madelung energies for Cu \& O are negative,
these changes largely cancel for
$\Delta\Phi_M$ and thus $\Delta$ (Fig. \ref{Fig:1}e) is largely unaffected.

Modifications in $\Delta$ allow us to examine the effect of the 
dopant on the exchange energy $J$ (Fig. \ref{Fig:1}f). 
Reflecting the spatial variations of $\Delta$, $J$
is suppressed by up to 20 \% near the dopant. 
Allocating dopant excess charge on
either on Cu, on O, or distributed on both result in slightly different
values of $J$ (Fig. \ref{Fig:1}g) but the overall
suppression seems rather immune to the way in which charge
is distributed.  

To test this in a non-perturbative way ED
studies of three-hole Cu$_2$O$_7$ (Fig. 2a) and Cu$_2$O$_8$ (Figs. 2b \& 2c) clusters were employed to
determine changes to the ZRS parameters\cite{EskesPRB1991}.  
Including Cu 3d$_{x^2-y^2}$, O 2p$_x$ and 
2p$_y$ orbitals, the Hamiltonian is $H = \sum_{i,\sigma} H_{i,\sigma}$ where: 
\begin{eqnarray}\label{Eq:Helec}\nonumber
H_{i,\sigma}&=&\epsilon_{d}^i d^\dagger_{i,\sigma}d_{i,\sigma}
             + \sum_{\delta} \epsilon_p^{i,\delta} p^\dagger_{i,\delta,\sigma}p_{i,\delta,\sigma}  \\ \nonumber 
            &+&\sum_{\delta} t_{pd}^{i,\delta} [d^\dagger_{i,\sigma}p_{i,\delta,\sigma} + h.c.]
             + \sum_{\delta,\delta'} t_{pp}^{\delta,\delta'} p^\dagger_{i,\delta,\sigma}p_{i,\delta',\sigma} \\
            &+&U_{dd}\hat{n}_{i,\uparrow}\hat{n}_{i,\downarrow} 
             + U_{pp}\sum_{\delta}\hat{n}_{i,\delta,\uparrow}\hat{n}_{i,\delta,\downarrow} 
\end{eqnarray}
where $\delta$ denotes the Cu-O basis vectors and $\hat{n}$ the number operator.   
From this 3-band model, an effective single-band Hubbard or $t-J$ model is
derived, with effective nearest $2t$ and next-nearest $2t^{\prime}$ neighbour hoppings
determined from the bonding/anti-bonding splitting of the ZRS, and $J$
derived from singlet-triplet splitting.
In terms of clusters Cu$_2$O$_7$ determines $t$ and $J$ as the ZRS 
involves a common bridging oxygen while Cu$_2$O$_8$ clusters 
yield $t^\prime$, $J^\prime$ via O-O hopping\cite{EskesPRB1991}.  
For the undoped lattice, $t = 383$, $t^{\prime} = -141$ meV and 
antiferromagnetic exchange couplings of $J = 218$, $J^{\prime} = 17$ meV are
characteristically obtained. $J$ may be fine-tuned by adjusting $t_{pd}$ and
$t_{pp}$.

\begin{figure}[t]
 \includegraphics[width=\columnwidth]{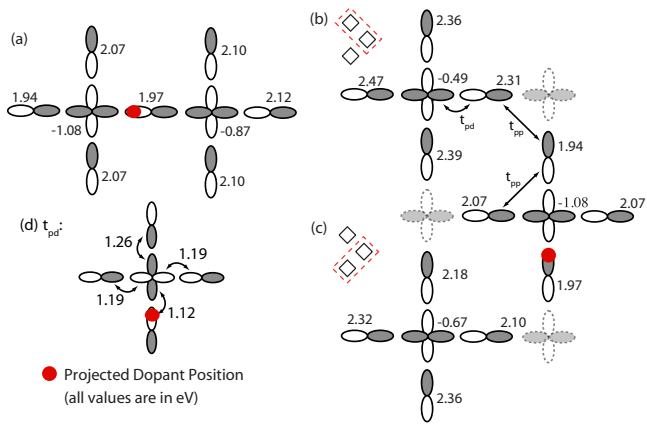}
 \caption{\label{Fig:SiteEnergies} 
(a) Cu$_2$O$_7$ cluster 
 used to extract $t$ and $J$.  (b) and (c) Two Cu$_2$O$_8$ 
 clusters shown in red outline used to extract $t^{\prime}$ and $J^{\prime}$.  The values for $\epsilon_{p,d}$ are indicated. (d) modifications to $t_{pd}$ 
 due to the dopant induced displacements.}
\end{figure}

The dopant is included by locally varying the on-site Cu and O energies 
$\epsilon_{d}$ and $\epsilon_{p}$, respectively, shown in
Fig. \ref{Fig:SiteEnergies}.
Here two Cu$_2$O$_8$ clusters are used which differ with respect to the location of the dopant, and  
the site energies have been determined from site modified
Madelung energies (Fig. 1), according to Eq. \ref{Eq:1}.
For the Cu$_2$O$_7$ cluster $t = 390$, $J = 216$ 
meV are obtained, while for the two Cu$_2$O$_8$ clusters (Figs. \ref{Fig:SiteEnergies}b \& \ref{Fig:SiteEnergies}c) 
we obtain $t^{\prime} = -242 (-189)$ meV and $J^{\prime} = 18 (17)$, respectively.  
Electrostatic modifications to $J$ and $J^{\prime}$ are slightly suppressed over
undoped cluster values, in general agreement with Madelung estimates, although the magnitude is smaller.
Importantly, we note that the symmetric placement of the dopant for the Cu$_2$O$_7$ cluster
gives only small changes to $t$, while for Cu$_2$O$_8$ the increases are noticably
larger, particularly for geometry Fig. \ref{Fig:SiteEnergies}b compared to 2c.
The asymmetric location of the dopant favors occupation of the oxygen ligand
orbitals in the plaquette containing the dopant,
giving larger modification of $t^\prime$.   
%It is an open issue of how to connect a large increase of $t^{\prime}$ to
%a tendency to locally favor $d_{x^2-y^2}$ pairing \cite{Pavarini}.

We have repeated these cluster calculations, including 
the modulations in $t_{pd}$ induced by the structural distortions, according to
the LDA displacements of Cu-O bond-distances\cite{HePRL2006, Harrison}, shown in Fig. 2d, and obtain 
$t = 374$, $J = 197$, 
$t^{\prime} = -231 (-180)$ and $J^{\prime} = 17 (16)$ meV. This tends to further
suppress $J$. Thus oxygen dopant's net effect is to mildly suppress $J$
and increase $t^{\prime}$, indicating that the dopant cannot
be viewed as only modifying site energies and increasing $J$ in 
downfolded single band models\cite{Maska}.
  
\begin{figure}[t]
 \includegraphics[width=8.0cm]{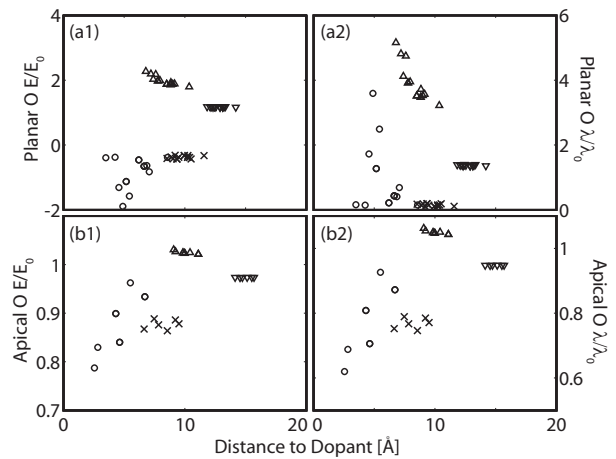}
 \caption{\label{Fig:3}
 (a1), (b1) Normalized local crystal fields at the planar and apical oxygen sites of the 
 doped lattice.  (a2), (b2) The corresponding electron phonon coupling strength 
 $\lambda \propto E^2$.  The data points follow the same key as figure 1.
 }
\end{figure}  

Since the Madelung energies are locally modified on the eV
scale, real space modifications may result in substantial changes to local
crystal fields. Ions vibrating
along the $c-$axis are sensitive to spatial gradients of the
Madelung energies, and 
Ref. \onlinecite{tpdPRB1999} has shown that these values of the local
field determines the overall strength of el-ph coupling at the
oxygen sites, controlling coupling to Raman active $A_{1g}$ planar and apical
oxygen vibrations, and out-of-phase planar $B_{1g}$ vibrations.
Therefore, we have examined the dopant's effect on the crystal field in
connection to local changes in el-ph coupling.

For the undoped crystal, c-axis crystal fields are
$E_{apex,plane} = 18.74, 1.16$ eV/$\AA$ for the apical, planar oxygen
sites, respectively, 
oriented as indicated in Fig. \ref{Fig:1}a. For the doped lattice, 
in Figs. \ref{Fig:3}a1, b1 we plot the local $c-$axis electric field strength
at the planar, apical oxygen sites, respectively, 
and the corresponding el-ph coupling strength $\lambda$ in
Figs. \ref{Fig:3}a2, b2\cite{tpdPRB1999}.  
One can see that the $E$-field strength is very sensitive to the local symmetry breaking 
introduced by the dopant, especially in the case of the planar oxygen atoms.  The presence 
of the dopant's bare charge in the otherwise positively charged SrO/BiO structure 
suppresses the field in the closest lying plane and the structural changes further modify 
the local fields.  The largest changes to the $E$-field occur in the plane whose field is 
oriented towards the dopant.  In this case the geometry is such that the dopant's bare charge 
increases the strength of the original field, driving $\lambda$ for the c-axis planar
oxygen modes up by a factor of 5. As it is well known that the $c-$axis phonons
give an attractive interaction in the $d_{x^2-y^2}$ channel, this suggests
that superconductivity may be locally
promoted by the dopant, in agreement with the assessment of
Ref. \onlinecite{Nunner}. The enhanced el-ph coupling may on the otherhand drive a tendency to locally bind a hole
to the lattice near the dopant location. This raises the possibility that LDOS 
modifications could be related to local polarons rather than pairing.
   
\begin{figure}[t]
 \includegraphics[width=8cm]{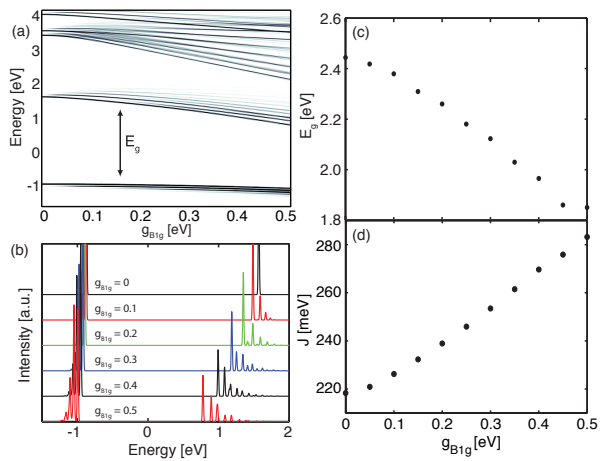}
 \caption{\label{Fig:4}
 (a) The electron addition and removal spectra for the CuO$_2$ cluster coupled to 
 c-axis oxygen vibrations as a function of el-ph coupling strength. The energy
 gap $E_g$ is indicated.
 (b) The electron addition (positive energy) and removal (negative) spectra as a function of energy for selected coupling 
 strengths. (c,d) $E_g$, $J$ as a function of el-ph coupling, respectively. }
\end{figure}  

To address this issue, we examine three-site polarons via ED
for an isolated CuO$_2$ Hubbard cluster coupled to c-axis oxygen vibrations:  
$H = H_{el} + H_{lat} + H_{el-ph}$, where $H_{el}$ is 
defined in Eq. \ref{Eq:Helec}, and 
$
H_{lat}=\sum_{\nu} \Omega_{\nu}\hat{n}^{\nu},~~~
H_{el-ph}= \sum_{\nu,\delta.\sigma} g_{\nu} (b_{\nu}^{\dagger}+b_{\nu}) e_{\nu}^{\delta}
p^\dagger_{\delta,\sigma}p_{\delta,\sigma}. $
Here $\hat{n}^{\nu}$ is the phonon number operator for branch $\nu$, $e_{\nu}^{\delta}$ is the polarization of 
the $\nu$-th quantized local displacement, and $g_{\nu} = eE\sqrt{\hbar/2M_O\Omega_{\nu}}$ sets
the el-ph coupling strength to mode $\nu$. We consider coupling to uniform out-of-phase $B_{1g}$ and in-phase $A_{1g}$ c-axis modes 
($\Omega=36$ and $55$ meV, respectively) coupled to the oxygen hole density by the local field $eE$. We have not included any local modifications to 
the phonon mode energies compared to the bulk.

To make contact to STM LDOS, the electron addition, removal spectra $A_{\pm}$, defined as 
\begin{equation}
\label{Eq:Addition}
A_{\pm}(\omega) = \sum_{i} |\bra{\Psi_i^{0,2}} c, c^{\dagger} \ket{\Psi_{gs}^1}|^2 \delta(\omega - E_i^{0,2} + E_{gs}^1),
\end{equation} 
where $\Psi^n_i$ denotes the $n$-hole states with energy $E^n_i$, are plotted in Figs. 4a \& 4b for different values of
el-ph coupling $g_{B_{1g}}$. A large number of phonon quanta have been used such that the results are independent of the 
number of phonons for the parameter region investigated.
As the coupling is increased, the spectral weight is gradually 
transferred into phonon side bands giving broader and suppressed spectral peaks.   
The corresponding energy gap $E_{g}$
between the first removal and addition states {\it decreases} with increasing $g_{B_{1g}}$
(Fig. \ref{Fig:4}c) as the effective charge transfer energy is reduced by the
gain in lattice energy. As a consequence, the value of $J$
(correspondingly determined from 2-hole Cu$_2$O$_7$ clusters with phonons)
increases substantially with increasing $g_{B_{1g}}$, as shown in Fig. 4d. 
The decrease of
$E_g$ and increase of $J$ due to local phonons will overwhelm the countering
effects from purely electrostatic considerations as the coupling to the lattice increases. 

To estimate the size of the effect for Bi-2212, $E_{plane}=1.16$ eV for the
undoped lattice yields $g_{B_{1g}}\sim 0.073$ eV, well into the large polaron
regime where side-bands are weak. Near the dopant however 
$g_{B_{1g}}$ is enhanced to $\sim 0.2$ eV, where side bands begin to develop spectral
weight in the removal/addition spectra and the charge gap is reduced (Figs. \ref{Fig:4}a-c).
At the same time, for this parameter regime, which is characteristic of multilayer cuprates, $J$
is increased by 20 meV. much greater than the small reduction determined from
electrostatic effects alone. Thus for realistic parameter regimes, c-axis phonons
act in concert with strong spatial variations of Madelung energies give an increase in $J$ as well. The dopant may then provide 
two coupled channels for $d-$wave pair enhancement, while causing suppressed
spectral features as a consequence of strong local e-ph coupling. This is
qualitatively what has been observed in experiments\cite{McElroy}. Such synergy between
phonons, polarons and antiferromagnetism has been already noticed in cluster quantum Monte Carlo studies\cite{DCA}.

In summary, we have investigated electrostatic modulations of local Madelung
energies arising from the presence of a dopant atom in Bi-2212 unit cells. While eV changes are found for the Madelung
energies for copper and oxygen, these changes largely cancel for the charge
transfer energy and give small local suppression to $J$. However the strong
local variations in Madelung energies are manifest in order 1
changes in el-ph coupling for $c-$axis oxygen modes. 
Using cluster studies, it was found that in combination electrostatic modifications and
coupling to the lattice yield broadened spectral features, reduced charge gap
energies $E_g$, and a sizable local increase of $J$, implying a 
strong local interplay between anti-ferromagnetism, polarons, and superconducting pairing. 
The amount of variation of the local charge gap can thus serve as an important
diagnostic for determining lattice coupling, electrostatic effects, and pairing. It is
an open question whether a link between these quantities can be made experimentally.

The authors would like to acknowledge valuable discussions with A. Balatsky,
J. C. Davis, W. Harrison, P. J. Hirschfeld, H.-P. Cheng, K. McElroy, 
B. Moritz, and J.-X. Zhu. This work was supported by NSERC.


\begin{thebibliography}{99}
\bibitem{Eisaki}
H. Eisaki {\it et al.}, Phys. Rev. B {\bf 69}, 064512 (2004).
\bibitem{Pavarini}
E. Pavarini {\it et al.}, Phys. Rev. Lett {\bf 87}, 047003 (2001).
\bibitem{OhtaPRB}
Y. Ohta, T. Tohyama and S. Maekawa, Phys. Rev. B {\bf 43}, 2968 (1991).
%\bibitem{STM-review}
%A. V. Balatsky, I. Vekhter, and J.-X. Zhu, Rev. Mod. Phys. {\bf 78}, 373 (2006).
\bibitem{McElroy}
K. McElroy {\it et al.}, Science {\bf 309}, 1048 (2005).
\bibitem{superlattice}
J. A. Slezak {\it et al}, Proc. Natl. Acad. Sci. USA {\bf 105}, 3203 (2008);
B. M. Andersen {\it et al.}, Phys. Rev. B {\bf 76}, 020507 (2007).
\bibitem{Nunner}
T. S. Nunner {\it et al.}, Phys. Rev. Lett. {\bf 95}, 177003 (2005); M. Mori
{\it et al.}, arXiv:0805.1281.
\bibitem{Maska}
J.-X. Zhu, cond-mat/0508646; M. M. Ma\'ska {\it et al}, Phys. Rev. Lett. {\bf 99}, 147006 (2007).
\bibitem{KovalevaPRB2004}
N. N. Kovaleva {\it et al},
%, A. V. Boris, R. Holden, C. Ulrich, B. Liang, C. T. Lin, B. Keimer, C. Bernhard, 
%J. L. Tallon, D. Munzar and A. M. Stoneham, 
Phys. Rev. B {\bf 69}, 54511 (2004). 
%\bibitem{Johnston}
%S. Johnston, F. Vernay, B. Moritz, T. P. Devereuax, Z.-X. Shen, N. Nagaosa and J. Zaanen 
%{\it To be published} 
\bibitem{params}
(in eV): 
$U_{dd} = 8.8$, $U_{pp} = 4.1$, $t_{pd} = 1.2$, $t_{pp} = 0.5$, $\epsilon_d = 0$ and 
$\epsilon_p = 2.92$ ,
where $U_{pp}$ and U$_{dd}$ are the on-site Coulomb repulsion for the O 2p and Cu 
3d$_{x^2-y^2}$ orbitals, $t_{pd}$ their overlap, and $t_{pp}$ is the O 2p-2p overlap integral. 
\bibitem{EskesPRB1993}
H. Eskes and J. H. Jefferson, Phys. Rev. B {\bf 48}, 9788 (1993).
\bibitem{TPDRH}
T. P. Devereaux and R. Hackl, Rev. Mod. Phys. {\bf 79}, 175 (2007).
\bibitem{HePRL2006}
Y. He {\it et al}, Phys. Rev. Lett. {\bf 96}, 197002 (2006).
\bibitem{EskesPRB1991}
H. Eskes and G. Sawatzky, Phys. Rev. B {\bf 43}, 119 (1991).
\bibitem{Harrison}
W. A. Harrison, {\it Elementary Electronic Structure}. World Scientific (2004).
%\bibitem{horsch}
%P. Horsch {\it et al.} Physica C {\bf 162}, 783 (1989).
%\bibitem{dagotto}
%E. Dagotto {\it et al.} Phys. Rev. Lett. {\bf 67}, 1918 (1991).
\bibitem{tpdPRB1999}
T. P. Devereaux {\it et al.}, Phys. Rev. B {\bf 59}, 14618 (1999).
\bibitem{DCA}
A. Macridin {\it et al.}, Phys. Rev. Lett. {\bf 97}, 056402 (2006).
\end{thebibliography}
\end{document}